\documentclass[a4paper,14pt]{extreport} 
\usepackage{dsfont}
\usepackage{amssymb}
\usepackage{amsfonts}
\usepackage{amsmath}
\usepackage{graphicx}
\usepackage{epsfig}
\usepackage{color}
\usepackage{amsthm}
\usepackage{fancybox,tikz}

\begin{document}

\newtheorem{definition}{Definition}
\newtheorem{proposition}{Proposition}
\newtheorem{example}{Example}
\newtheorem{theorem}{Theorem}
\newtheorem{lemma}{Lemma}
\newtheorem{corollary}{Corollary}
\newtheorem{remark}{Remark}

\begin{titlepage}
\begin{center}
\vspace{0.1in} {\Large Institute of Computer Science}
\par
\vspace{0.1in} {\Large Romanian Academy}
\par
\vspace*{1in} {\Huge\bf Spatial Dynamic Structures and Mobility in
Computation}
\par
\vspace{1.5in} {\Large Bogdan Aman}
\par
\vfill
\par
\vspace{0.1in} {\Large Supervisor}
\par
\vspace{0.1in} {\Large Prof. Dr. Gabriel Ciobanu}
\par
\vspace{0.5in} October 2009
\end{center}
\end{titlepage}

\begin{abstract}
Membrane computing is a well-established and successful research
field which belongs to the more general area of molecular computing.
Membrane computing aims at defining parallel and non-deterministic
computing models, called membrane systems or P Systems, which
abstract from the functioning and structure of the cell. A membrane
system consists of a spatial structure, a hierarchy of membranes
which do not intersect, with a distinguishable membrane called skin
surrounding all of them. A membrane without any other membranes
inside is elementary, while a non-elementary membrane is a composite
membrane. The membranes define demarcations between regions; for
each membrane there is a unique associated region. Since we have a
one-to-one correspondence, we sometimes use membrane instead of
region, and vice-versa. The space outside the skin membrane is
called the environment.

In this thesis we define and investigate variants of systems of
mobile membranes as models for molecular computing and as modelling
paradigms for biological systems. On one hand, we follow the
standard approach of research in membrane computing: defining a
notion of computation for systems of mobile membranes, and
investigating the computational power of such computing devices.
Specifically, we address issues concerning the power of operations
for modifying the membrane structure of a system of mobile membranes
by mobility: endocytosis (moving a membrane inside a neighbouring
membrane) and endocytosis (moving a membrane outside the membrane
where it is placed). On the other hand, we relate systems of mobile
membranes to process algebra (mobile ambients, timed mobile
ambients, $\pi$-calculus, brane calculus) by providing some
encodings and adding some concepts inspired from process algebra in
the framework of mobile membrane computing.
\end{abstract}

\chapter*{Acknowledgements}

I thanks to my supervisor Dr. Gabriel Ciobanu for providing this
opportunity, and for his constant support and advice throughout this
PhD.

I thanks the anonymous referees for their comments and suggestions
which helped improve the quality of the papers in which parts of
this thesis were published.

I would also like to thanks my examiners: Dr. Gheorghe P\u aun, Dr.
Dorel Lucanu, Mitic\u a Craus, Matteo Cavaliere, for their careful
reading of the thesis and their helpful comments.

I am grateful to all my colleagues from the group of Formal Methods
Laboratory, Institute of Computer Science, Romania Academy, Ia\c si
Branch for their suggestions and encouragements.

\chapter*{Contents}

\begin{tabular}{l}
$~$\hspace{70ex}$~$\\
1 Introduction \dotfill 2\\
\qquad 1.1 Context \dotfill 2\\
\qquad 1.2 Motivation \dotfill 3\\
\qquad 1.3 Outline \dotfill 4\\
2 Preliminaries \dotfill 6\\
\qquad 2.1 Alphabets, strings and languages \dotfill 6\\
\qquad 2.2 Chomsky Grammars \dotfill 7\\
\qquad 2.3 L Systems \dotfill 9\\
\qquad 2.4 Matrix Grammars \dotfill 10\\
\qquad 2.5 Register Machines \dotfill 12\\
\qquad 2.6 Process Algebra \dotfill 12\\
\qquad \qquad 2.6.1 $\pi$-calculus \dotfill 12\\
\qquad \qquad 2.6.2 Safe Mobile Ambients \dotfill 13\\
\qquad \qquad 2.6.3 Brane Calculi \dotfill 16\\
\qquad 2.7 P Systems \dotfill 17\\
\qquad \qquad 2.7.1 Transition P Systems \dotfill 17\\
\qquad \qquad 2.7.2 P Systems with Active Membranes
\dotfill 18\\
3 Mobile Membranes \dotfill 24\\
\qquad 3.1 Simple Mobile Membranes \dotfill 25\\
\qquad \qquad 3.1.1 Definition \dotfill 25\\
\qquad \qquad 3.1.2 Computational Power \dotfill 26\\
\qquad 3.2 Enhanced Mobile Membranes \dotfill 28\\
\qquad \qquad 3.2.1 Motivation \dotfill 28\\
\qquad \qquad 3.2.2 Definition \dotfill 31\\
\qquad \qquad 3.2.3 Modelling power \dotfill 32\\
\qquad \qquad 3.2.4 Computational Power \dotfill 35\\
\qquad 3.3 Mutual Mobile Membranes \dotfill 37\\
\qquad \qquad 3.3.1 Motivation \dotfill 37\\
\qquad \qquad 3.3.2 Definition \dotfill 39\\
\qquad \qquad 3.3.3 Computational Power \dotfill 39\\
\qquad 3.4 Conclusions \dotfill 45\\
\end{tabular}

\begin{tabular}{l}
$~$\hspace{67ex}$~$\\
4 Mobile Membranes with Objects on Surface \dotfill 46\\
\qquad 4.1 Mutual Membranes with Objects on Surface \dotfill 47\\
\qquad \qquad 4.1.1 Definition \dotfill 47\\
\qquad \qquad 4.1.2 Computational Power \dotfill 48\\
\qquad \qquad 4.1.4 Relationship of PEP Calculus and Mutual\\
\qquad\qquad Membranes with Objects on Surface\dotfill 53\\
\qquad 4.2 Conclusions \dotfill 58\\
5 Membrane Systems and Process Algebra \dotfill 59\\
\qquad 5.1 Mobile Membranes and Mobile Ambients \dotfill 60\\
\qquad \qquad 5.1.2 Relationship of Safe Mobile Ambients with\\
\qquad\qquad Mutual Mobile Membranes \dotfill 60\\
\qquad \qquad 5.1.3 Decidability Results \dotfill 80\\
\qquad 5.2 Time in Mobile Membranes and Mobile Ambients \dotfill 89\\
\qquad \qquad 5.2.1 Time in Membrane Computing \dotfill 89\\
\qquad \qquad 5.2.2 Mobile Membranes with Timers \dotfill 90\\
\qquad \qquad 5.2.3 Mobile Membranes With and Without Timers \dotfill 92\\
\qquad \qquad 5.2.4 Mobile Membranes with Timers and\\
\qquad \qquad Timed Mobile Ambients \dotfill 95\\
\qquad 5.3 Typed Membrane Systems \dotfill 99\\
\qquad \qquad 5.3.1 Typed Membrane Systems \dotfill 99\\
\qquad \qquad 5.3.2 Typed $\pi$-calculus \dotfill 108\\
\qquad 5.4 Conclusions \dotfill 111\\
6 Conclusion \dotfill 113\\
\qquad 6.1 Contributions \dotfill 113\\
\qquad 6.2 Other Contributions \dotfill 115\\
\end{tabular}

\newpage

\chapter*{Preliminaries}
\label{chapter:preliminaries}

This chapter contains some basic notions of formal language theory,
as well as automata. For further information about these topics the
reader is referred to the monographs
\cite{Dassow90,Garey79,Hopcroft79,Rozenberg97} and to specific
papers cited in the next sections.

\newpage

\chapter*{Mobile Membranes}

In this chapter we define the systems of simple, enhanced and mutual
mobile membranes and study their modelling and computational power.

The {\it systems of simple mobile membranes} are a variant of P
systems with active membranes having none of the features like
polarizations, label change, division of non-elementary membranes,
priorities, or cooperative rules. The additional feature considered
instead are the operations of {\it endocytosis} and {\it
exocytosis}: moving a membrane inside a neighbouring membrane, or
outside the membrane where it is placed. However, these operations
are slightly different in the papers introducing them: in
\cite{Csuhaj-Varju08} one object is specified in each membrane
involved in the operation, while in \cite{KrishnaPaun05} one object
is mentioned only in the moving membrane. Another variant of P
systems with mobile membranes is mobile P systems \cite{Petre99-02}
having rules inspired from mobile ambients \cite{Cardelli98}. Turing
completeness is obtained by using nine membranes together with the
operations of endocytosis and exocytosis \cite{KrishnaPaun05}. Using
also some contextual evolution rules (together with endocytosis and
exocytosis), in \cite{Krishna05} it is proven that four mobile
membranes are enough to get the power of a Turing machine, while in
\cite{LNBI09} we decrease the number of membranes to three. In order
to simplify the presentation, we use {\it systems of simple mobile
membranes} instead of P systems with mobile membranes.

The {\it systems of enhanced mobile membranes} are a variant of
membrane systems which we proposed in \cite{FBTC07} for describing
some biological mechanisms of the immune system. The operations
governing the mobility of the systems of enhanced mobile membranes
are endocytosis (endo), exocytosis (exo), forced endocytosis (fendo)
and forced exocytosis (fexo). The computational power of the systems
of enhanced mobile membranes using these four operations was studied
in \cite{Krishna08} where it is proved that twelve membranes can
provide the computational universality, while in \cite{LNBI09} we
improved the result by reducing the number of membranes to nine. It
is worth to note that unlike the previous results, the rewriting of
object by means of context-free rules is not used in any of the
results (proofs).

Following our approach from \cite{SYNASC09} we define {\it systems
of mutual mobile membranes} representing a variant of systems of
simple mobile membranes in which the endocytosis and the exocytosis
work whenever the involved membranes ``agree'' on the movement; this
agreement is described by using dual objects $a$ and $\overline{a}$
in the involved membranes. The operations governing the mobility of
the systems of mutual mobile membranes are mutual endocytosis
(mutual endo), and mutual exocytosis (mutual~exo). It is enough to
consider the biologically inspired operations of mutual endocytosis
and mutual exocytosis and three membranes (compartments) to get the
full computational power of a Turing machine \cite{UC09}. Three
represents the minimum number of membranes in order to discuss
properly about the movement provided by endocytosis and exocytosis:
we work with two membranes inside a skin membrane.

\newpage

\chapter*{Mobile Membranes with Objects on Surface}

Membrane systems \cite{Paun00,Paun02} and brane calculus
\cite{Cardelli04} have been inspired from the structure and the
functioning of the living cell. Although these models start from the
same observation, they are build having in mind different goals:
membrane systems investigate formally the computational nature and
power of various features of membranes, while the brane calculus is
capable to give a faithful and intuitive representation of the
biological reality. In \cite{CardelliPaun05} the initiators of these
two formalisms describe the goals they had in mind: ``While membrane
computing is a branch of natural computing which tries to abstract
computing models, in the Turing sense, from the structure and the
functioning of the cell, making use especially of automata,
language, and complexity theoretic tools, brane calculi pay more
attention to the fidelity to the biological reality, have as a
primary target systems biology, and use especially the framework of
process~algebra.''

In \cite{CBM08} we define a new class of systems of mobile
membranes, namely the {\it systems of mutual membranes with objects
on surface}. The inspiration to add objects on membrane and to use
the biologically inspired rules pino/exo/phago comes from
\cite{Brijder08,CardelliPaun05,Cavaliere07,Krishna07,Krishna09}. The
novelty comes from the fact that we use objects and co-objects in
phago and exo rules in order to illustrate the fact that both
involved membranes agree on the movement. We investigate in
\cite{NaCo09} the computational power of systems of mutual membranes
with objects on surface controlled by pairs of rules: pino/exo or
phago/exo, proving that they are universal with a small number of
membranes. Similar rules are used by another formalism called brane
calculus \cite{Cardelli04}. We compare in \cite{CBM08} the systems
of mutual membranes with objects on surface with brane calculus, and
encode a fragment of brane calculus into the newly defined class of
systemms of mobile membranes. Even brane calculus have an
interleaving semantic and membrane systems have a parallel one, by
performing this translation we show that the difference between the
two models is not significant.

\newpage

\chapter*{Membrane Systems and Process Algebra}

The membrane systems \cite{Paun00,Paun02} and the mobile ambients
\cite{Cardelli98} have similar structures and common concepts. Both
have a hierarchical structure representing locations, and are used
to model various aspects of biological systems. The mobile ambients
are suitable to represent the movement of ambients through ambients
and the communication which takes place inside the boundaries of
ambients. Membrane systems are suitable to represent the movement of
objects and membranes through membranes. We consider these new
computing models used in describing various biological phenomena
\cite{Cardelli04,Ciobanu06}, and encode the ambients into membrane
systems \cite{MeCBIC06}. We present such an encoding, and use it to
describe the sodium-potassium exchange pump \cite{BIOSYSTEMS08}. We
provide an operational correspondence between the safe ambients and
their encodings, as well as various related properties of the
membrane systems \cite{BIOSYSTEMS08}.

In \cite{WMC07} we investigate the problem of reaching a
configuration from another configuration in a special class of
systems of mobile membranes. We prove that the reachability can be
decided by reducing it to the reachability problem of a version of
pure and public ambient calculus without the capability {\sf open}.

A feature of current membrane systems is the fact that objects and
membranes are persistent. However, this is not quite true in the
real world. In fact, cells and intracellular proteins have a
well-defined lifetime. Inspired from these biological facts, we
define in \cite{CompMod09} a model of mobile membranes in which each
membrane and each object has a timer attached representing their
lifetime. This new feature is inspired from biology where cells and
intracellular proteins have a well-defined lifetime. In order to
simulate the passage of time we use rules of the form $a^{\Delta t}
\rightarrow a^{\Delta (t-1)}$ and $[~]_i^{\Delta t} \rightarrow
[~]_i^{\Delta (t-1)}$ for the objects and membranes which are not
involved in other rules. If the timer of an objects reaches $0$ then
this object is consumed by applying a rule of the form $a^{\Delta 0}
\rightarrow \lambda$, while if the timer of a membrane $i$ reaches
$0$ then the membrane is dissolved by applying a rule of the form
$[~]_i^{\Delta 0} \rightarrow [\delta]_i^{\Delta 0}$. After
dissolving a membrane, all objects and membranes previously present
in it become elements of the immediately upper membrane, while the
rules of the dissolved membrane are~removed. Some results show that
mutual mobile membranes with and without timers have the same
computational power. Since we have defined an extension with time
for mobile ambients in \cite{CSR07,ICTAC07,FORTE08}, and one for
mobile membrane in \cite{CompMod09}, we study the relationship
between these two extensions: timed safe mobile ambients are encoded
into mutual mobile membranes with timers.

Membrane systems \cite{Paun00,Paun02} are known to be Turing
complete and are often used to model biological systems and their
evolution. In order to increase the modelling power of this
formalism, we define in \cite{WMC09} a typed version, which leads in
turn to a decrease of computational power. We enrich the
symport/antiport membrane systems with a {\it type discipline} which
allow to guarantee the soundness of reduction rules with respect to
some relevant properties of the biological systems. The key
technical tools we use are type inference and principal typing
\cite{Wells02}, i.e. we associate to each reduction rule the minimal
set of conditions that must be satisfied in order to assure that
applying this rule to a correct membrane system, then we get a
correct membrane system as well. The type system for membrane
systems with symport/antiport rules is (up to our knowledge) the
first attempt to control the evolution of membrane systems using
typing rules. The presentation of the typed sodium-potassium pump is
an example how to introduce and use types in membrane systems. The
$\pi$-calculus typed pump is presented in order to see what is the
desired modelling power we want to have in membrane systems by
introducing a type system. Type descriptions of biological inspired
formalisms, along with the type inference algorithm can also be
found in \cite{MeCBIC08,Fages06}. Other static techniques have been
applied to biological systems, such as Control Flow Analysis
\cite{Bodei09,Nielson04,Nielson07} and Abstract
Interpretation~\cite{Fages08}.

\vspace{-1.5ex}

\newpage

\chapter*{Conclusions}

\section*{Contributions}

Membrane computing is a well-established and successful research
field which belongs to the more general area of molecular computing.
Membrane computing aims at defining parallel and non-deterministic
computing models, called membrane systems or P Systems, which
abstract from the functioning and structure of the cell. Since the
introduction of this model, many variants have been proposed and the
literature on the subject is now rapidly growing. There are two
standard ways of investigating membrane systems: considering their
computational power in comparison with the classical notion of
Turing computability, or considering their efficiency in solving
algorithmically hard problems, like NP-problems, in a polynomial
time.

In this respect, we defined new classes of membrane systems which
are powerful, mostly equivalent to Turing machines, and for which we
established links with process algebra.

The {\it systems of simple mobile membranes} are a variant of P
systems with active membranes having none of the features like
polarizations, label change, division of non-elementary membranes,
priorities, or cooperative rules. The additional feature considered
instead are the operations of {\it endocytosis} and {\it
exocytosis}: moving a membrane inside a neighbouring membrane, or
outside the membrane where it is placed. In \cite{LNBI09} we proved
that three mobile membranes are enough to get the power of a Turing
machine.

The {\it systems of enhanced mobile membranes} are a variant of
systems of simple mobile membranes that we proposed in \cite{FBTC07}
for describing some biological mechanisms of the immune system. The
operations governing the mobility of the systems of enhanced mobile
membranes are endocytosis (endo), exocytosis (exo), enhanced
endocytosis (fendo) and enhanced exocytosis (fexo). In \cite{LNBI09}
we studied the computational power of the systems of nine enhanced
mobile membranes using these four operations.

Following our approach from \cite{SYNASC09} we defined {\it systems
of mutual mobile membranes} representing a variant of systems of
simple mobile membranes in which the endocytosis and exocytosis work
whenever the involved membranes ``agree'' on the movement; this
agreement is described by using dual objects $a$ and $\overline{a}$
in the involved membranes. The operations governing the mobility of
the systems of mutual mobile membranes are mutual endocytosis
(mutual endo), and mutual exocytosis (mutual~exo). It is enough to
consider the biologically inspired operations of mutual endocytosis
and mutual exocytosis and three membranes (compartments) to get the
full computational power of a Turing machine \cite{UC09}.

In \cite{CBM08} we defined a new class of systems of mobile
membranes, namely the {\it systems of mutual membranes with objects
on surface}. The rules of this class are biologically inspired,
namely pino/exo/phago rules. The novelty comes from the fact that we
used objects and co-objects in phago and exo rules in order to
illustrate the fact that both involved membranes agree on the
movement. We investigated in \cite{NaCo09} the computational power
for systems of mutual membranes with objects on surface controlled
by pairs of rules: pino/exo or phago/exo, proving that they are
universal with a small number of membranes. Similar rules are used
by another formalism called brane calculus \cite{Cardelli04}. In
\cite{CBM08} we defined an operational semantic for systems of
membranes with objects on surface and compare the systems of mutual
membranes with objects on surface with brane calculus, and encoded a
fragment of brane calculus into the newly defined class of systems
of mobile membranes.

In \cite{MeCBIC06} we encoded the mobile ambients into the membrane
systems and provided an operational semantic for membrane systems.
We presented such an encoding, and used it to describe the
sodium-potassium exchange pump \cite{BIOSYSTEMS08}. We provided an
operational correspondence between the safe ambients and their
encodings, as well as various related properties of the membrane
systems \cite{BIOSYSTEMS08}.

In \cite{WMC07} we investigated the problem of reaching a
configuration from another configuration for a subclass of systems
of mobile membranes, and proved that the reachability can be decided
by reducing it to the reachability problem of a version of pure and
public ambient calculus without the capability {\sf open}.

In \cite{SYNASC09-02,CompMod09} new classes of membranes are
defined: timers are assigned to each membrane and each object. This
new feature is inspired from biology where cells and intracellular
proteins have a well-defined lifetime. In order to simulate the
passage of time we use rules of the form $a^{\Delta t} \rightarrow
a^{\Delta (t-1)}$ and $[~]_i^{\Delta t} \rightarrow [~]_i^{\Delta
(t-1)}$ for the objects and membranes which are not involved in
other rules. If the timer of an objects reaches $0$ then this object
is consumed by applying a rule of the form $a^{\Delta 0} \rightarrow
\lambda$, while if the timer of a membrane $i$ reaches $0$ then the
membrane is dissolved by applying a rule of the form $[~]_i^{\Delta
0} \rightarrow [\delta]_i^{\Delta 0}$. After dissolving a membrane,
all objects and membranes previously present in it become elements
of the membrane containing it, while the rules of the dissolved
membrane are~removed.

By adding timers to objects and membranes into a system of mutual
mobile membranes, we do not obtain a more powerful formalism.
According to \cite{CompMod09} we have that systems of mutual mobile
membranes with timers and systems of mutual mobile membranes without
timers have the same power. Since an extension with time for mobile
ambients already exists \cite{CSR07,ICTAC07,FORTE08}, and one for
systems of mobile membranes is presented in \cite{CompMod09}, the
relationship between these two extensions is studied: timed safe
mobile ambients are encoded into systems of mutual mobile membranes
with timers \cite{CompMod09}.

In \cite{WMC09} we enriched the symport/antiport membrane systems
with a {\it type discipline} which allows to guarantee the soundness
of reduction rules with respect to some relevant properties of the
biological systems. The key technical tools we used are type
inference and principal typing, i.e. we associate to each reduction
rule the minimal set of conditions which must be satisfied in order
to assure that applying this rule to a correct membrane system, then
we get a correct membrane system as well. The type system for
membrane systems with symport/antiport rules is (up to our
knowledge) the first attempt to control the evolution of membrane
systems using typing rules. The presentation of the typed
sodium-potassium pump is an example how to introduce and use types
in membrane systems. The $\pi$-calculus typed pump is presented in
order to see what is the desired modelling power we want to have in
membrane systems by introducing a type system.

\section*{Other Contributions}

We also focused on other formalisms which are characterized by
spatial dynamic structures and mobility, namely mobile ambients and
the calculus of looping sequences. In what follows we present the
work done in this direction.

In \cite{CSR07} we extended mobile ambients with timers and
proximities, in order to get a clear notion of location and
mobility. Timers define timeouts for various resources, making them
available only for a determined period of time; we add timers to
ambients and capabilities. We presented an example how the new model
is working. The coordination of the ambients in time and space is
given by assigning specific values to timers, and by a set of
coordination rules.

In \cite{ICTAC07} we added timers to communication channels,
capabilities and ambients, and used a typing system for
communication. The passage of time is given by a discrete global
time progress function. We proved that structural congruence and
passage of time do not interfere with the typing system. Moreover,
once well-typed, an ambient remains well-typed. A timed extension of
the cab protocol illustrates how the new formalism is working.

In \cite{FORTE08} we added timers to capabilities and ambients, and
provided an operational semantics of the new calculus. Certain
results are related to the passage of time, and some new behavioural
equivalences over timed mobile ambients are defined. Timeout for
network communication (TTL) can be naturally modelled by the time
constraints over capabilities and ambients. The new formalism can be
used to describe network protocols; Simple Network Management
Protocol (SNMP) may implement its own strategy for timeout and
retransmission in TCP/IP.

In \cite{MeCBIC08} we enriched the calculus of looping sequences, a
formalism for describing evolution of biological systems by means of
term rewriting rules, with type disciplines to guarantee the
soundness of reduction rules with respect to interesting biological
properties.

\begin{minipage}{15cm}
\begin{tabular}{lr}
\includegraphics[width=1.75in]{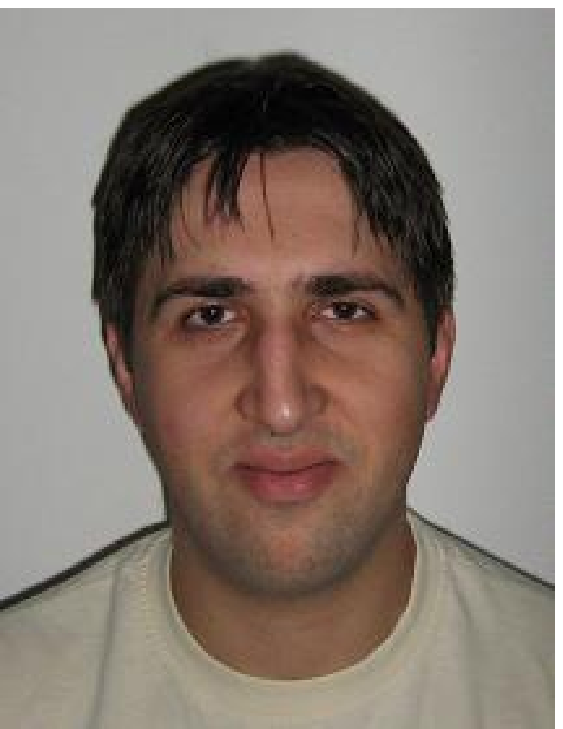}&
\begin{minipage}{10cm} \textbf{Bogdan Aman}, born at 26 June 1982,
Botosani, has graduated ``Al.I.Cuza'' University of Ia\c si, Faculty
of Mathematics, in 2007. He is a Ph.D. student under the supervision
of Dr. Gabriel Ciobanu at the Romanian Academy (Ia\c si Branch),
Institute of Computer Science. His main research fields are membrane
computing, computational modelling for systems biology, process
algebra, and other theoretical aspects of computer science.

\end{minipage}\\
\end{tabular}
\end{minipage}

\end{document}